\documentclass{article}

\usepackage{float}

\usepackage{arxiv}
\usepackage[utf8]{inputenc} 
\usepackage[T1]{fontenc}    
\usepackage{hyperref}       
\usepackage{url}            
\usepackage{booktabs}       
\usepackage{amsfonts}       
\usepackage{nicefrac}       
\usepackage{microtype}      
\usepackage{lipsum}		
\usepackage{graphicx}
\usepackage{natbib}
\usepackage{doi}
\usepackage{gensymb}

\usepackage{xcolor}
\usepackage{subfig}

\usepackage{tikz} 
\usetikzlibrary{shapes.geometric, arrows.meta, positioning, calc}

\tikzstyle{startstop} = [circle, draw=black]
\tikzstyle{process} = [rectangle, minimum width=4cm, minimum height=1cm, text centered, text width=3.75cm, draw=black]
\tikzstyle{box} = [rectangle, draw, minimum width=30mm, minimum height=9mm, align=center]
\tikzstyle{decision} = [diamond, minimum width=3cm, minimum height=1cm, text centered, draw=black]
\tikzstyle{arrow} = [thick,->,>=stealth]

\usepackage{amsmath}
\title{Global Dataset of Solar Power Plants: Multidimensional Integration and Analysis}

\author{ \href{https://orcid.org/0009-0002-1387-8982}{\includegraphics[scale=0.06]{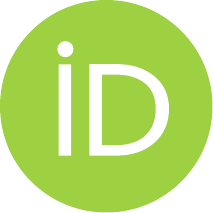}\hspace{1mm}Anibal Mantilla-Guerra} \\
	FIGEMPA\\
	Central University of Ecuador\\
	Quito, Ecuador\\
	\texttt{armantilla@uce.edu.ec} \\
	\And
	\href{https://orcid.org/0000-0001-6715-191X}{\includegraphics[scale=0.06]{orcid.pdf}\hspace{1mm}Christian Mejia-Escobar} \\
	FIGEMPA\\
	Central University of Ecuador\\
	Quito, Ecuador \\
	\texttt{cimejia@uce.edu.ec} \\
   	\And
	\href{https://orcid.org/0000-0003-4762-6927}{\includegraphics[scale=0.06]{orcid.pdf}\hspace{1mm}Jorge Azorin-Lopez} \\
	Department of Computer Technology\\
	University of Alicante\\
	Alicante, Spain \\
	\texttt{jazorin@ua.es} \\
   	\And
    \href{https://orcid.org/0000-0002-7798-3055}{\includegraphics[scale=0.06]{orcid.pdf}\hspace{1mm}Jose Garcia-Rodriguez} \\
	Department of Computer Technology\\
	University of Alicante\\
	Alicante, Spain \\
	\texttt{jgarcia@dtic.ua.es} \\
   	\And
	\href{https://orcid.org/0000-0001-6890-7563}{\includegraphics[scale=0.06]{orcid.pdf}\hspace{1mm}Byron Fernando Tarco} \\
	Department of Geology\\
	Central University of Ecuador\\
	Quito, Ecuador \\
	\texttt{bftarco@uce.edu.ec} \\
   	\And
	\href{https://orcid.org/0000-0000-0000-0000}{\includegraphics[scale=0.06]{orcid.pdf}\hspace{1mm}Karen Santamaria} \\
	Department of Geology\\
	Central University of Ecuador\\
	Quito, Ecuador \\
	\texttt{knsantamaria@uce.edu.ec} \\
}

\renewcommand{\shorttitle}{\textit{arXiv} Template}

\hypersetup{
pdftitle={A template for the arxiv style},
pdfsubject={q-bio.NC, q-bio.QM},
pdfauthor={David S.~Hippocampus, Elias D.~Striatum},
pdfkeywords={First keyword, Second keyword, More},
}

\begin{document}
\maketitle

\begin{abstract}

The use of clean energy is a global trend, with solar photovoltaic plants
serving as a cornerstone of this energy transition. To support this rapid growth, optimize energy utilization, and enable a wide range of applications and services, it is essential to have access to more sophisticated and detailed solar data. Specifically, existing datasets lack integration, contain significant gaps, and have limited geographic coverage. In contrast, this study proposes a reliable, standardized, and multidimensional dataset with a global scope. Through a reproducible methodology and automated processes, we have successfully collected, generated, and combined 27 attributes of geographic, topographic, logistical, climate, and power nature, which are critical for the study of photovoltaic plants worldwide. Based on descriptive statistical
analysis of the 58,978 records comprising the compiled dataset, the raw data have been transformed into valuable information for the energy sector. This demonstrates the utility of this product as a source of knowledge discovery, publicly available to the academic and professional communities.
\end{abstract}


\section{Introduction}
\label{sec:intro}
Humanity has never been so dependent on energy.
It is the driving force behind today's world and underpins every aspect of modern life. 
However, rising energy demand and excessive reliance on fossil fuels, along with their environmental impacts, represent some of the most pressing challenges of our time.
The search for cleaner and more sustainable alternatives has propelled solar energy to become a cornerstone of the global energy transition
\cite{Lijiang2025}\cite{Mekhilef2011}.

Among solar energy systems, photovoltaic (PV) plants are the most widespread. Their principle is simple yet powerful: converting solar radiation—a free, inexhaustible resource found everywhere on the planet—into renewable, environmentally friendly electricity \cite{Betak2020}.
PV technology has established itself as a mature and viable solution due to its rapid industrial development and the steady decline in manufacturing costs \cite{Fang2018}.
Given their widespread expansion, it is essential to have comprehensive information on solar plants around the world.
This is a key factor in its strategic deployment and the diversification of the energy matrix.

Although there is a wide range of information available on solar energy installations worldwide, the existing datasets have significant limitations: They are inconsistent, incomplete, have partial geographical coverage, and are scattered across isolated public and private sources.
Furthermore, none of them combines in a single product the various attributes that most broadly define PV plants and optimize the use of solar energy.
This lack of integrated, comprehensive, and global data represents a clear scientific gap, which is the reason for the development of this project.

Our specific goal has been to create a high-quality dataset on photovoltaic plants worldwide by combining geographic, topographic, climatic, logistic, and power attributes.
These factors are fundamental and indispensable; they not only describe the existing solar infrastructure but also offer numerous benefits and a wide range of applications, such as solar energy analysis, strategic planning, reducing technical and financial uncertainty, optimizing the location and design of future installations, guiding decision-making in both the public and private sectors, scientific research, and training artificial intelligence models.

The implementation of this proposal poses technical and methodological challenges.
The massive growth in solar power plants has also increased the complexity of collecting, organizing, cleaning, and integrating data into a single dataset.
The approach we use to identify the most representative variables is based on a cross-validation technique that combines scientific relevance and technical feasibility, overcoming the individual limitations of each source or technological platform related to the solar industry.
The subsequent steps in creating the dataset constitute a fully reproducible methodology, supported by computational tools that significantly reduce the time and effort required for its development.

On the one hand, a Geographic Information System (GIS) allows us to create and integrate various layers of information about topography and land suitability through the geospatial processing of Digital Elevation Models (DEMs) \cite{Goncalves2025}.
On the other hand, the Python programming language allows us to implement automated processes for extracting and downloading data by querying the APIs of information sources, as well as to generate data that is essential but whose availability is limited or nonexistent.
For example, additional information on solar radiation, climate, and logistics variables such as area and proximity to road infrastructure \cite{Sharma2025}.

Combining these tools has made it possible to automate a large part of the workflow.
As a result, a total of 58,978 records were obtained from countries around the world, each containing 27 key variables related to geography, topography, logistics, climate, and power.
This dataset of solar photovoltaic plants meets the highest standards for sophistication, comprehensiveness, and size compared to similar products, while also meeting requirements for consistency, reliability, geographic diversity, and updating.

This manuscript is structured as follows: Section \ref{sec:intro} provides an overview of the project. Section \ref{sec:methods} offers a detailed description of the proposed dataset and the workflow designed for its creation. Section \ref{sec:stats} is devoted to the statistical analysis of the dataset’s attributes to gain insights. Finally, Section \ref{sec:conclusion} presents the conclusions and future work.

\section{Methodology}
\label{sec:methods}
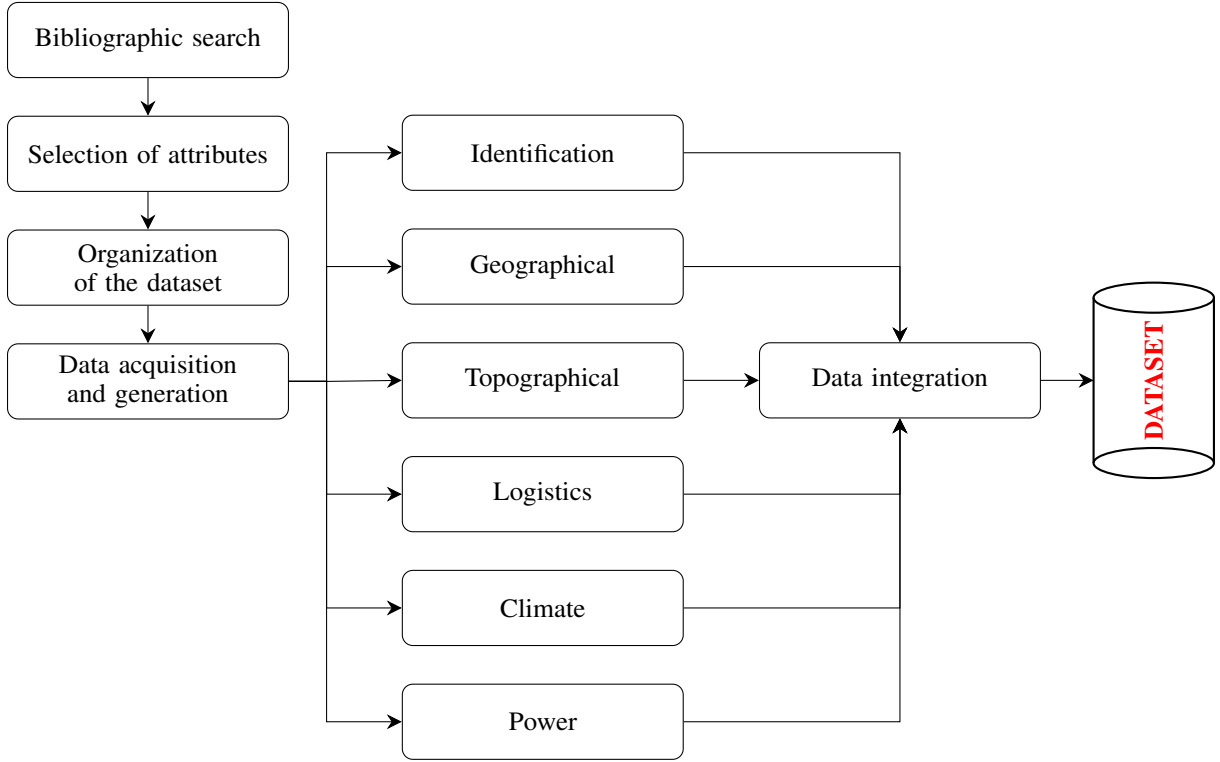
\begin{figure}[!htb]
    \centering
\begin{tikzpicture}[
    block/.style={rectangle, draw, text width=3.5cm, text centered, rounded corners, minimum height=1cm, minimum width=3.5cm},
    line/.style={draw, -{Stealth[length=2mm, width=2mm]}},
    cylinder/.style={cylinder, draw, shape border rotate=90, aspect=0.4, minimum height=3cm, minimum width=1.5cm, fill=white}
]

\node[block] (busqueda) at (0,0) {Bibliographic search};
\node[block, below=0.5cm of busqueda] (seleccion) {Selection of attributes};
\node[block, below=0.5cm of seleccion] (organizacion) {Organization of the dataset};
\node[block, below=0.5cm of organizacion] (adquisicion) {Data acquisition and generation};

\node[block, right=1.5cm of busqueda.east, yshift=-1.5cm] (identificacion) {Identification};
\node[block, below=0.5cm of identificacion] (geograficos) {Geographical};
\node[block, below=0.5cm of geograficos] (topograficos) {Topographical};
\node[block, below=0.5cm of topograficos] (logisticos) {Logistics};
\node[block, below=0.5cm of logisticos] (climaticos) {Climate};
\node[block, below=0.5cm of climaticos] (energeticos) {Power};

\node[block, right=1cm of topograficos.east] (integracion) {Data integration};

\coordinate (dataset-center) at ([xshift=1.5cm]integracion.east);
\coordinate (dataset-left) at ([xshift=-0.8cm]dataset-center); 
\draw[thick] (dataset-center) ++(0,-1.1) ellipse (0.8 and 0.2); 
\draw[thick] (dataset-center) ++(0,1.1) ellipse (0.8 and 0.2);  
\draw[thick] (dataset-center) ++(-0.8,-1.1) -- ++(0,2.2);         
\draw[thick] (dataset-center) ++(0.8,-1.1) -- ++(0,2.2);          
\node[font=\bfseries\color{red}, rotate=90] at (dataset-center) {DATASET};

\path[line] (busqueda) -- (seleccion);
\path[line] (seleccion) -- (organizacion);
\path[line] (organizacion) -- (adquisicion);

\path[line] (adquisicion.east) -- ++(0.5,0) |- (identificacion.west);
\path[line] (adquisicion.east) -- ++(0.5,0) |- (geograficos.west);
\path[line] (adquisicion.east) -- ++(0.5,0) -- (topograficos.west);
\path[line] (adquisicion.east) -- ++(0.5,0) |- (logisticos.west);
\path[line] (adquisicion.east) -- ++(0.5,0) |- (climaticos.west);
\path[line] (adquisicion.east) -- ++(0.5,0) |- (energeticos.west);

\path[line] (identificacion.east) -| (integracion.north);
\path[line] (geograficos.east) -| (integracion.north);
\path[line] (topograficos.east) -- (integracion.west);
\path[line] (logisticos.east) -| (integracion.south);
\path[line] (climaticos.east) -| (integracion.south);
\path[line] (energeticos.east) -| (integracion.south);

\path[line] (integracion.east) -- (dataset-left);

\end{tikzpicture}

    \caption{Methodology used to create the multidimensional dataset of solar plants.}
    \label{fig:flujograma}
\end{figure}

The project focuses on creating a high-quality dataset on solar photovoltaic plants worldwide.
The methodology used to create the dataset is illustrated in Figure\ref{fig:flujograma}.
This methodology consists of five main stages: a literature review to identify information and data sources; the definition of attributes; the structuring of the dataset; and the collection and generation of data covering various aspects of solar power plants. Finally, the integration and cleaning of this data into a single dataset are performed.
 
The final product successfully combines various types of variables into a single, coherent dataset that integrates both quantitative and qualitative attributes, with its quality rooted in the methodological rigor of the data collection, cleaning, labeling, and validation phases.
The importance of creating accurate and up-to-date registries of solar installations in various regions is highlighted, particularly in emerging markets where the potential for solar energy is enormous.
The following describes each of the steps required to obtain the final product.

\subsection{Bibliographic search}
The goal is to create a dataset that is theoretically robust and technically feasible.
By considering this dual approach—what science tells us is necessary and what technical reality tells us is feasible to extract—we ensure that the resulting dataset not only contains the scientifically necessary variables but is also technically feasible to implement and replicate using existing global data infrastructures.
Consequently, the initial phase of the project involved a thorough search for high-impact research with scientific credibility, as well as the main existing data sources on solar power plants.

First, AI-assisted research tools such as \textit{STORM} \cite{storm2024} and \textit{Consensus} \cite{consensus2024} were used, which enabled the filtering of existing literature based on keyword searches such as: ``dataset, solar panels, photovoltaic, power, and AI''.
The results obtained excluded informal websites, opinion pieces, and studies on other types of renewable energy.
The purpose of this broad-ranging search was to identify common patterns in reputable research and prioritize those studies that the scientific community considers critical.
Only articles published in the last 10 years in journals indexed in databases such as \textit{Elsevier} and \textit{WoS}, as well as academic repositories such as \textit{Google Scholar}, were included.

Second, the search for available data sources focused on the Web.
Since its emergence in the 1990s, the volume of available information has covered every subject \cite{berners-lee1994} \cite{gillies}.
On this global platform, information about solar power plants, energy resources, and geospatial conditions has grown exponentially.
Our review of the literature on the datasets most commonly used in climate studies and solar projects is summarized in Table \ref{tab:bases_radiacion}.

\begin{table}[!htb]
\centering
\caption{Summary of solar data sources identified during the literature review.}
\label{tab:bases_radiacion}
\resizebox{\textwidth}{!}{
\begin{tabular}{clcccccc}
\toprule
\textbf{\#} & \textbf{Organization} & \textbf{Access} & \ \textbf{Main variables} & \textbf{Period} & \textbf{Applications} \\
\midrule
1 & NASA POWER \cite{nasa_power} & Free & GHI, DNI, DHI, temp., wind, clouds & 1983--current & Preliminaries, AI \\
2 & PVGIS \cite{pvgis} & Free & GHI, DNI, diffuse, albedo, temp. & 2005--current & PV design, simulations \\
3 & Global Solar Atlas \cite{globalsolaratlas} & Free/Paid & GHI, DNI, diffuse, temp., production & >20 years & Solar plants \\
4 & SolarGIS \cite{solargis} & Paid & GHI, DNI, weather, dirt, deterioration & >20 years & Engineering, digital twins \\
5 & Meteonorm \cite{meteonorm} & Paid (demo) & GHI, DNI, weather & Sintetic & Filling gaps, simulation \\
6 & ERA5 (Copernicus) \cite{era5} & Free & GHI, clouds, wind, temp. & 1979--current & Predictive models, AI \\
7 & MERRA-2 (NASA) \cite{merra2} & Free & GHI, clouds, aerosols & 1980--current & AI, weather \\
8 & CERES (NASA) \cite{ceres_dataset} & Free & Incoming/outgoing radiation, clouds & 2000--current & Energy balance \\
9 & HelioClim (HC3/HC4) \cite{helioclim} & Free/Paid & GHI, DNI & 2004--current & PV homogeneous series \\
10 & BSRN \cite{bsrn} & Free & Surface radiation & 1992--current & High precision \\
11 & SURFRAD (NOAA) \cite{surfrad} & Free & Radiation, weather & 1995--current & Accurate measurement \\

\bottomrule
\end{tabular}}
\end{table}

The identified solar data sources include well-known organizations in the fields of solar energy, meteorology, and climate modeling.
Depending on their applications, three main categories can be distinguished: 1) satellite data sources (NASA, MERRA-2, CERES, HelioClim) and climate reconstructions or reanalyses (ERA5); 2) ground-based measurement networks (BSRN, SURFRAD); and 3) simulation and estimation tools for solar engineering (PVGIS, Global Solar Atlas, SolarGIS, and Meteonorm).

Although there is a close relationship between these organizations, they differ in the aspects they measure, the variables they consider, and the level of accuracy of the information they provide.
Given the diverse nature of energy data sources, none can be considered comprehensive on its own.
Each has a specific purpose: some offer excellent radiation resolution but lack environmental data, while others provide global climate data but have poor solar accuracy.

Combining these different sources tends to yield better results than using just one.
Although it is possible to extract a variety of information, combining it involves dealing with issues such as inconsistencies between global satellite data and point measurements on the ground, spatial resolution, temporal standardization, and differences across geographic regions, among others.
Therefore, not only is it justified to create a dataset that brings together the most representative variables in order to eliminate individual biases from each source and provide high-fidelity data for decision-making, but the uniqueness of this work lies in the simultaneous consolidation of geographic, topographic, logistical, climatic, and power variables into a single global repository, an integration that is rarely found in the current literature.

\subsection{Selection of attributes}

Selecting the most appropriate variables to build a robust dataset is a task that researchers undertake based on their specific needs \cite{Guerrero2024}.
In this case, the selection was made through a two-phase systematic review.
On the one hand, a comparative analysis of the documents obtained from the literature review revealed that various variables have been proposed in the literature over time.
For this reason, priority was given to those with the greatest academic consensus, relevance, and frequency of occurrence in the sources consulted.
On the other hand, the most widely recognized data sources were analyzed to determine which variables are actually available.
By cross-referencing both analysis, critical attributes—both numerical and categorical—were identified and grouped according to technical criteria into the following key areas: identification, geographic location, topography, logistics, climate, and power.

The inclusion of identifying attributes is justified because they serve as the basis for uniquely and clearly distinguishing each solar plant within the dataset.
Geographic attributes make it possible to understand the spatial variability of solar resources, visualize and verify them, and integrate data from other sources into each plant.
Topographic attributes describe the characteristics of the terrain that determine exposure to solar radiation and the suitability of a site for the installation of solar power plants, directly influencing energy generation potential, facility design, and project feasibility.
Logistics attributes describe the infrastructure and accessibility conditions of the surrounding area that influence the feasibility and construction of solar power plants.
Climate variables are essential for assessing the availability of energy resources and the conditions that affect the efficiency of solar modules.
Finally, power-related attributes make it possible to correlate a site’s potential with its actual generation capacity.

\subsection{Organization of the dataset}

Table \ref{tab:dataset} lists the 27 attributes that compose the proposed dataset. For each attribute, the table specifies the assigned name, category, variable type, unit of measurement, and a brief description of its meaning.

\begin{table}[!htb]
	\caption{Structure of the dataset of solar plants and main characteristics of its attributes.}
	\centering
\scalebox{0.9}{    
	\begin{tabular}{lllll}
		\toprule
		\textbf{Attribute} & \textbf{Category} & \textbf{Type} & \textbf{Units} & \textbf{Description} \\
		\midrule
		Code & Identification & Qualitative  & - & Unique alphanumeric identifier for the plant \\
        Name &  &  Qualitative  & - & Denomination of the solar power plant \\
        \hline
		Country   & Geographical  & Qualitative  & - & National territorial entity where the plant is located \\
        Latitude &  & Quantitative & Grades & Geographic coordinates associated with parallels \\
        Longitude &  & Quantitative & Grades & Geographic coordinate associated with meridians \\
        \hline
        Elevation & Topographical & Quantitative & $m$ & Vertical height of the solar plant \\   
        Slope &  & Quantitative & Grades & Angle of inclination of the terrain \\ 
        Aspect &  & Quantitative & Grades & Direction of the slope \\
        Curvature &  & Quantitative & - & The shape of the terrain \\
        T\_slope &  & Qualitative & - & Classification of slope \\
        T\_aspect &  & Qualitative & - & Classification of aspect  \\
        T\_curvature &  & Qualitative & - & Classification of curvature \\
        \hline
        Distance to road & Logistics & Quantitative & $m$ & Distance from the solar plant to the nearest road \\
        Area &  & Quantitative & $m^{2}$ & Total extension of the solar plant\\        
        Size &  & Qualitative & -& Classification of area \\
        \hline
        Temperature &  Climate & Quantitative & $\circ C$ & Average ambient air temperature at the site \\
        Humidity &  & Quantitative & \% & Water vapor content in the atmosphere \\
        Wind speed &  & Quantitative & $m/s$ & Airflow velocity at the site \\  
        Wind direction &  & Quantitative & Grades & The cardinal point from which the wind blows \\
        DT\_wind &  & Qualitative & - & Classification of wind direction \\
        \hline
   		Operational status & Power & Qualitative  & - & Current phase of the project lifecycle \\        
        Solar aptitude &  & Quantitative & - & A site suitability score for a solar power plant \\
        T\_aptitude &  & Qualitative & - & Classification of solar aptitude \\
        Optimal tilt &  & Quantitative & Grades & Ideal solar panel angle \\ 
        Global irradiation &  & Quantitative & $kWh/m^{2}$ & Energy incident on a surface over a period of time \\        
        Capacity &  & Quantitative & $kWp$ & Total installed power \\ 
        PV potential &  & Quantitative & $kWh/kWp$ & Amount of energy generated per kWp \\
        \bottomrule
	\end{tabular}
}    
	\label{tab:dataset}
\end{table}

An innovative feature of the project is the calculation of the area and solar aptitude index for each PV plant. These attributes are considered extremely valuable and are not typically found in general information sources.
In addition, to facilitate the analysis and interpretation of certain numerical variables (area, slope, aspect, curvature, wind, and solar aptitude), their respective categorical versions were created based on these variables (size, T\_slope, T\_aspect, T\_curvature, DT\_wind, and T\_aptitude).

Once the variables that form the dataset have been specified, the dataset is organized into a tabular format.
In this structure, each column corresponds to a predefined attribute, while each row represents a registered solar plant instance.
The next step is to obtain the data needed to complete that structure, which involves collecting the actual values for each variable.
In some cases, these values can be obtained directly from available data sources, while in others they must be generated during the dataset creation process.

\subsection{Data acquisition and generation}

This section explains where the data was obtained and how it was collected.
Effective data collection depends on the integration of various datasets from different sources, such as government agencies and private organizations.
Table \ref{tab:sources} presents the main sources for viewing and downloading the data for the variables included in the dataset, as well as the method used to obtain that data.

\begin{table}[!htb]
	\caption{Data sources for the variables included in the dataset of solar power plants.}
	\centering
	\begin{tabular}{llll}
        \toprule
		\textbf{Source of information} & \textbf{Data format} & \textbf{Variable} & \textbf{Method} \\
		\midrule
        Global Energy Monitor (GEM) & Data downloaded in & Name & Direct extraction \\
         &  CSV format & Country & Direct extraction \\
         &  & Latitude & Direct extraction \\        
         &  & Longitude & Direct extraction \\
         &  & Operational status & Direct extraction \\
         &  & Capacity & Direct extraction \\    
         \hline
        Global Solar Atlas (GSA) & Raster file in & Elevation & Direct extraction \\
         &  GeoTiff format  &  Optimal tilt & Direct extraction \\
         &  & Global irradiation & Direct extraction \\
         &  & PV potencial & Direct extraction \\
         \hline
        NASA POWER & CSV, raster and & Temperature & Direct extraction \\
        \& ERA5 (Copernicus) & NetCDF (matrix) & Humidity & Direct extraction \\
         &  & Wind speed & Direct extraction \\
         &  & Wind direction & Direct extraction \\
         
         \hline
        Our own generation & Digital Elevation & Code & Manual assignment \\
         & Models (DEMs),  & Slope & Geospatial calculation \\
         & raster layers, & Aspect & Geospatial calculation \\
         & vector layers,  & Curvature & Geospatial calculation \\
         & road layers from & T\_slope & Classification \\
         & GEOFABRIK y & T\_aspect & Classification \\
         & OpenStreetMap & T\_curvature & Classification \\
         & (OSM) & Distance to road & Geometric calculation \\
         &  & Area & Geometric calculation \\
         &  & Size & Classification \\ 
         &  & DT\_wind & Classification \\
         &  & Solar aptitude & Geospatial calculation \\
         &  & T\_aptitude & Classification \\
        \bottomrule
	\end{tabular}
	\label{tab:sources}
\end{table}

All of the sources mentioned are open access and useful for research, planning, and decision-making.
The following section describes the data collection process by category: identification, geographic location, terrain characteristics, logistics, climate, and operational characteristics of solar power plants.

\subsubsection{Identification data}

The identification data includes the \textit{code} and the \textit{name} of the solar power plant.
Each plant is assigned a unique alphanumeric code generated as part of the dataset creation process.
It consists of three elements: a unique integer value (ObjectID) that the GIS automatically assigns to each record (row) in the table, the first three letters of the plant’s country of origin, and the first letter of the plant’s size category (P, M, or G).
This design provides a structured and informative identifier that makes it easier to organize, search, and analyze the dataset.
For example, the code “16056-ECU-P” identifies a small-scale solar power plant in Ecuador called “Intiyana Solar Farm”.

\begin{figure}[!htb]
	\centering
	\fbox{\includegraphics[width=0.75\linewidth]{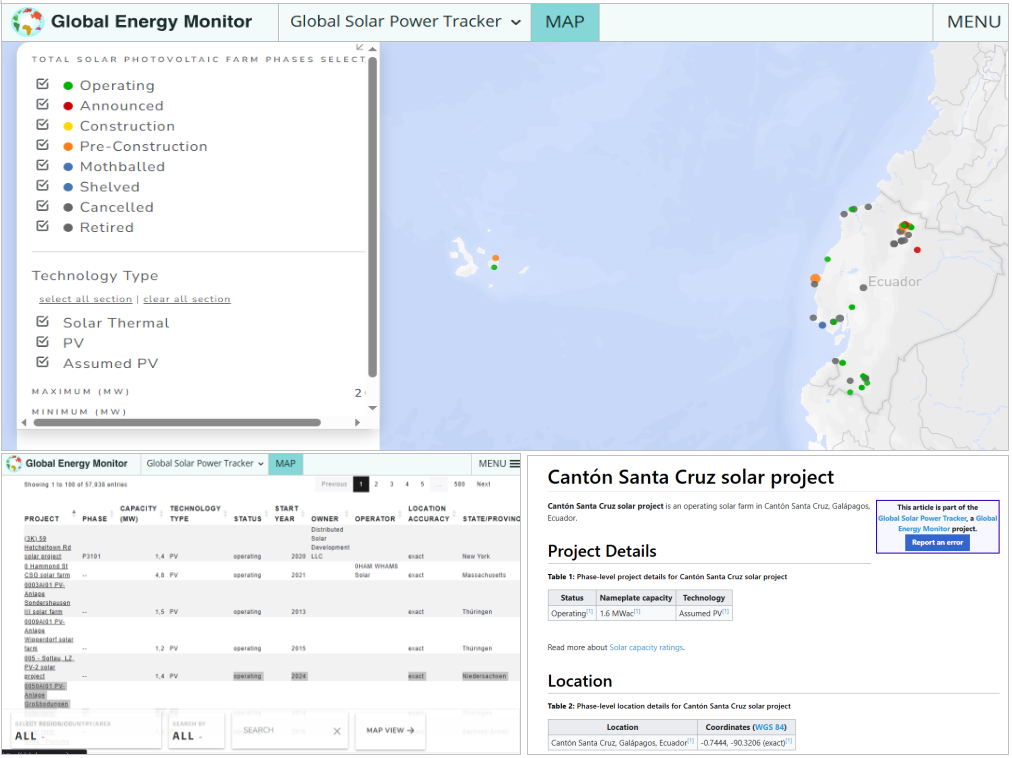}}
	\caption{Global Energy Monitor (GEM) website.}
	\label{fig:gem}
\end{figure}

The name of each solar plant corresponds to the official designation of the PV project on the website of Global Energy Monitor (GEM), a global reference for the world’s energy infrastructure, covering both fossil fuels and renewable energy \cite{globalenergymonitor}.
Through its platform and interactive tools, users can view the geographic distribution of energy infrastructure and other key data.
The website provides an interactive world map (Figure \ref{fig:gem}) that allows users to view high-resolution solar data for any location on the planet.
GIS data layers can be downloaded for analysis in specialized software programs such as QGIS or ArcGIS.

\begin{figure}[!htb]
	\centering
	\fbox{\includegraphics[width=0.75\linewidth]{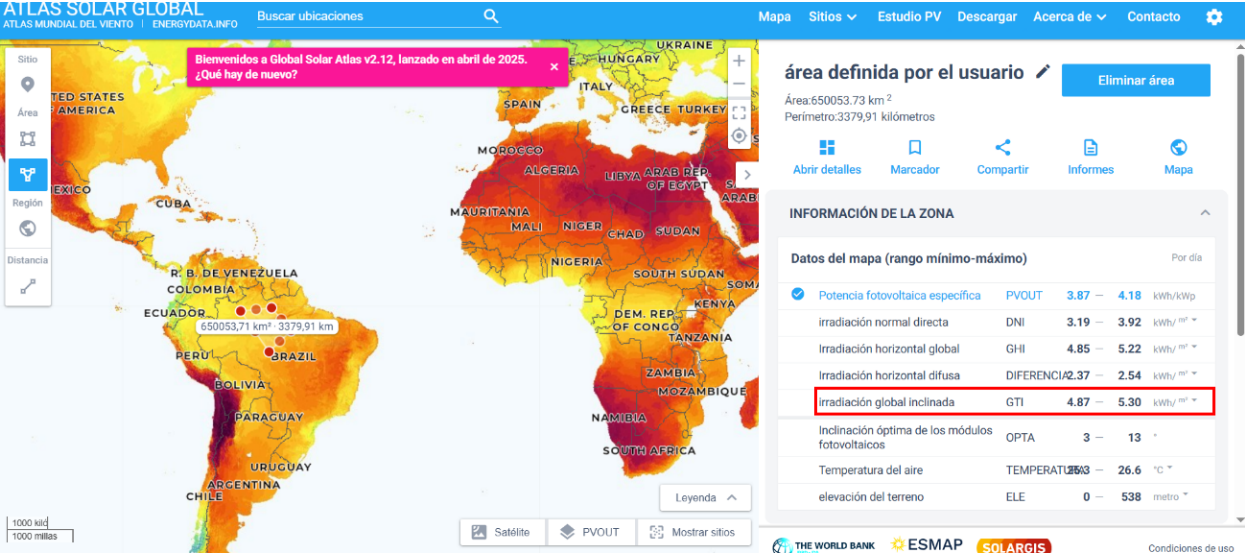}}
	\caption{Global Solar Atlas (GSA) website.}
	\label{fig:atlassolar}
\end{figure}

Although the vast majority of solar plant names come from GEM, in specific cases, data from the Global Solar Atlas (GSA) web platform was used (Figure \ref{fig:atlassolar}).
This is because some names in GEM were encoded in such a way that they are unreadable.
GSA is a free online tool developed by the World Bank Group and the company Solargis.
It is a key resource for energy planning, research, and, in particular, for identifying the optimal locations for solar power plants worldwide \cite{globalsolaratlas}.

\subsubsection{Geographical data}

\begin{figure}[!htb]
    \centering
    \includegraphics[width=0.8\linewidth]{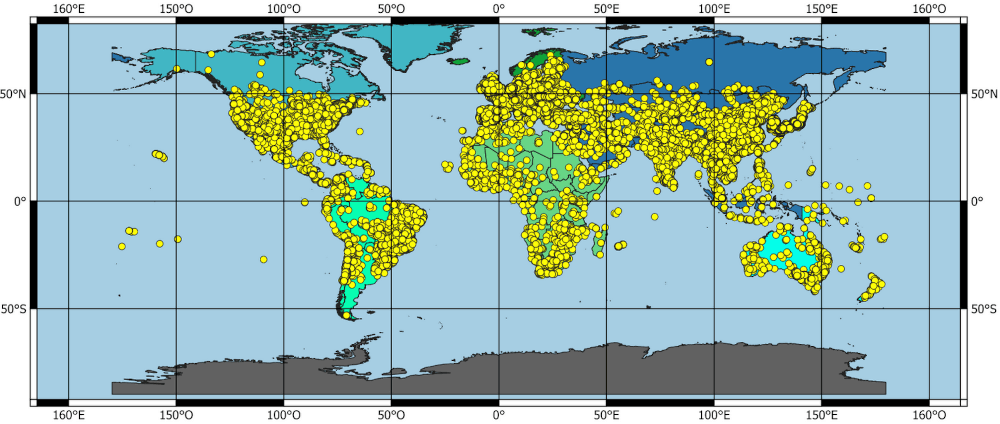}
    \caption{Global location map of the solar plants in the dataset.}
    \label{fig:plantsmap}
\end{figure}

The data on \textit{country}, \textit{latitude}, and \textit{longitude} for each solar plant were extracted directly from the GEM portal, 
under the “Global Solar Power Tracker” option, which is a tracker that collects information on energy projects around the world \cite{gem_globalsolarpowertracker_2026}.
Here, users are asked to fill out a basic form (name, institutional email address, and purpose of use), which allows them to download the solar plant inventory.
Once access was approved, the system provided the dataset files in standard CSV, Shapefile (SHP), and GeoJSON formats.
These files were integrated into the GIS by adding vector layers (SHP and GeoJSON) and delimited text files (CSV), defining the latitude and longitude columns, and assigning the WGS 84 (EPSG:4326) spatial reference system to ensure their correct location.
Next, the georeferencing and spatial consistency of the points were verified, generating the resulting layer shown in Figure \ref{fig:plantsmap}.

\subsubsection{Topographical data}

These factors are critical for determining the optimal location for solar power plants.
These variables are mathematically derived from a Digital Elevation Model (DEM).
A DEM is a three-dimensional representation of the Earth's surface in the form of a regular pixel-based grid, where each cell or pixel contains an elevation value.
As a result, raster maps of \textit{elevation}, shading, \textit{slope}, \textit{aspect}, and \textit{curvature} were generated, all of which retain this same matrix format.

The DEM is downloaded directly from the GSA platform in GeoTIFF format, which is compatible with GIS environments and has a spatial resolution of 9 arc-seconds appropriate for the study’s objectives.
The coverage area is organized by continent, so we will use South America as an example.
Before using the DEM, it is essential to prepare it properly to perform the required spatial analyses by following these steps in ArcMap:

\begin{enumerate}
    \item Reproject to a coordinate system in meters since the DEMs were originally in WGS84 (degrees).
    The “South America Albers Equal Area Conic” projection was chosen, which not only allows for the standardization of units of measurement for distance calculations but as an equivalent projection also preserves the proportionality of areas across the entire continent.
    This ensures that the orientation of the slopes and the surface area of the land identified as suitable for the installation of solar panels are geographically reliable.
    \item Verify the integrity of the DEM by identifying the following spatial inconsistencies:
    
\begin{itemize}
    \item Irregular patterns such as artificially flat surfaces, lines or blocks with no data, and abrupt changes in elevation that do not correspond to the actual terrain.
    \item Check the value assigned to "NoData" using the Source tab of the raster. If there is no suitable mask, generate an explicit mask using tools such as "Set Null" or the raster editor, thereby ensuring that the valid areas are correctly delineated.
    \item Correct artificial depressions or peaks using the Fill tool, as these errors can disrupt the data flow and produce incorrect results (Spatial Analyst → Hydrology → Fill). This is shown in Figure \ref{fig:lidar}.

\begin{figure}[!htb]
	\centering
	\fbox{\includegraphics[width=0.5\linewidth]{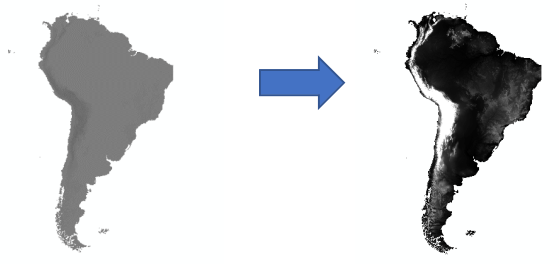}}
	\caption{Fill correction.}
	\label{fig:lidar}
\end{figure}

    \item Apply moderate smoothing if necessary: Filter (Low Pass) and Focal Statistics (mean, median).
    \item Adjust the Z-factor to ensure consistency between horizontal and vertical units; typically, set it to 1 when working in meters. The Z-factor affects the calculation of slope, curvature, aspect, and shading.
\end{itemize}
    \item Finally, the DEM should be clipped to the study area to improve performance and ensure spatial consistency with the other layers (Data Management → Raster → Raster Processing → Clip). Ensure consistent resolution across multiple layers: Resample if necessary.

\end{enumerate}

\begin{figure}[!htb]
	\centering
\subfloat[Shadow map.]{\fbox{\includegraphics[width=3.6cm, height=4.75cm]{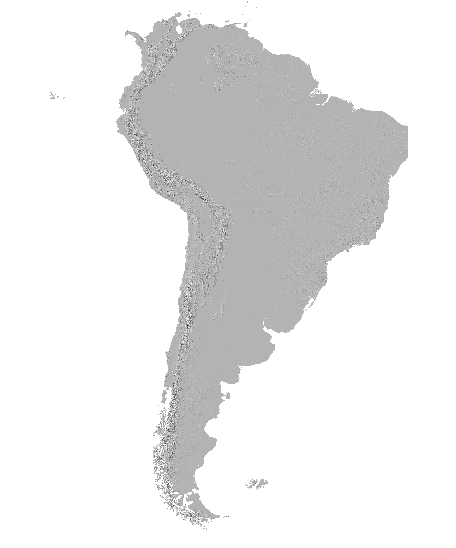}}}
\subfloat[Slope map.]{\fbox{\includegraphics[width=3.6cm, height=4.75cm]{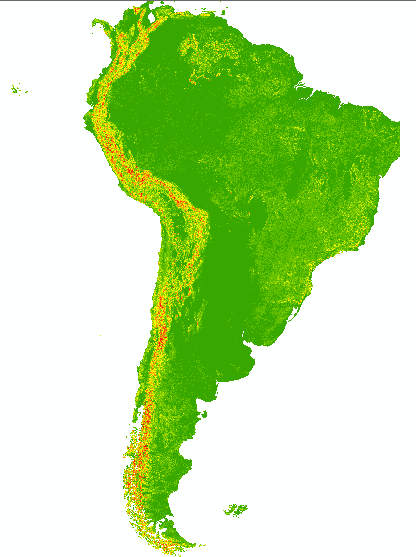}}}
\subfloat[Curvature map.]{\fbox{\includegraphics[width=3.6cm, height=4.75cm]{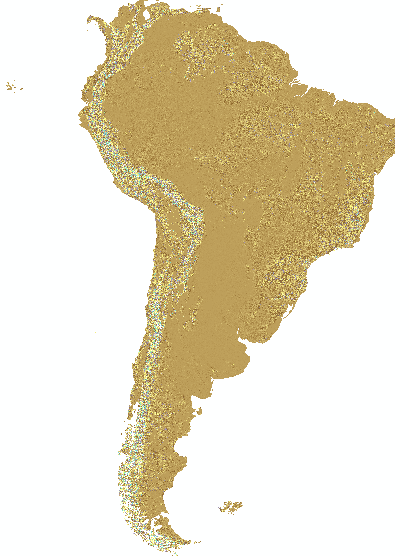}}}
\subfloat[Aspect map.]{\fbox{\includegraphics[width=3.6cm, height=4.75cm]{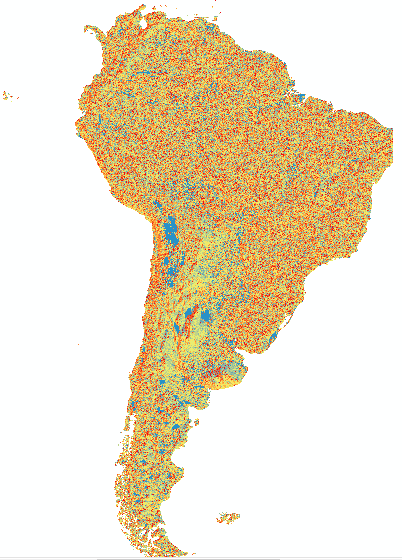}}}
   
	\caption{Raster maps of topographic variables derived from the DEM of South America.}
	\label{fig:maps}
\end{figure}

Based on the cleaned, continuous, and correctly projected DEM, derived maps are automatically generated with the same pixel resolution of 100 x 100 m.
These maps are generated in ArcGIS using the "Terrain Analysis" tool from the “Spatial Analyst” module. The resulting raster maps are shown in Figure \ref{fig:maps}.

The shadow map highlights the contours of the terrain, such as mountains and valleys, using standard lighting parameters (azimuth 315° and elevation 45°).
The slope map shows the incline of the terrain expressed in degrees or as a percentage, which is essential for land suitability analysis.
The curvature map depicts convex surfaces associated with ridges or watersheds (positive values), concave surfaces associated with areas of accumulation or valleys (negative values), and flat or transitional areas (values close to zero).
The aspect map determines the orientation of slopes on a scale from 0° to 360°, assigning a value of –1 to completely flat areas; this is important because exposure to sunlight and wind directly affects the performance of a solar power plant.

The above maps are essential tools for identifying areas with PV potential, as they allow for an accurate assessment of the topographical conditions that directly influence solar energy capture.
Each map provides additional information, and when combined, they help identify optimal, safe, and efficient locations for installing solar panels.
The maps described represent variables that, in their original form, are numerical.
However, they were converted into categorical variables to facilitate their interpretation and statistical analysis.
This allows continuous values to be grouped into representative classes, which enables the identification of patterns and comparisons between solar power plants.

The qualitative variable named \textit{T\_slope} is generated from the slope classification, which ranges from 0–2\degree (“Flat or nearly flat”), 2–5\degree (“Smooth”), 5–10\degree (“Moderate”), 10–15\degree (“Steep”), 15–30\degree (“Very steep”) up to $\ge$ 30\degree (“Steep or abrupt”).
The slope ranges used are consistent with FAO recommendations for classifying slopes in soil and land-use studies \cite{FAO1976}\cite{Verheye2009}, allowing the slope of the terrain to be correlated with factors such as erosion, stability, and land-use potential, thereby facilitating their application in solar suitability studies and land-use planning.

The classification of aspect generates the \textit{T\_aspect} variable, which has 8 cardinal directions (N, NE, E, SE, S, SW, W, NW), with a special category for flat slopes (“Flat”).
This categorization is based on the convention of dividing the 360° of a circle into 45° sectors, which simplifies the interpretation of terrain orientation; it is widely used in geomorphology, hydrology, and terrain analysis in GIS.

Finally, the terrain curvature classification (\textit{T\_curvature}) consists of three categories: flat or intermediate surfaces (values close to zero, between $-0.01$ and 0.01), concave surfaces or valleys (negative values), and convex surfaces or ridges (positive values).
This interpretation of curvature is part of the methodology described by Wilson and Gallant \cite{inbook} for consistently distinguishing between flat areas, depressions, and elevations.

\subsubsection{Logistics data}

Data on \textit{distance to road}, \textit{area}, and \textit{size} were included to describe the accessibility and extent of the solar plants.
The generation of this data through integration and calculation processes constitutes the main contribution of this work, as it provides information that is not available in existing datasets or sources.

Firstly, the distance from the solar plants to the main road network was calculated as follows:

\begin{enumerate}
    \item Download road maps of South America in a vector file (shapefile) from the GEOFABRIK repository, a service that facilitates the extraction of OSM road layers at the regional level \cite{geofabrik}.

    \item Filtering these layers in QGIS using SQL queries, selecting only first- and second-class roads.

    \item Merging the resulting vector files into a single layer or shapefile, which is then integrated into the solar plant inventory.

    \item A topological correction is performed using the “Correct Geometries” algorithm applied to the road network to correct digitization errors (discontinuities, gaps, etc.).

    \item Reproject both layers (roads and plants) to the “South America Albers Equal Area (EPSG:102033)” reference system, ensuring a metric projection at the continental level.

    \item Run the “Near” tool in ArcMap to identify the nearest neighbor between the two layers, generating the final vector layer that contains the distance in meters.
\end{enumerate}
    
\begin{figure}[!htb]
	\centering
\subfloat[Manual distance to road measurement.]{\fbox{\includegraphics[width=7cm, height=4.5cm]{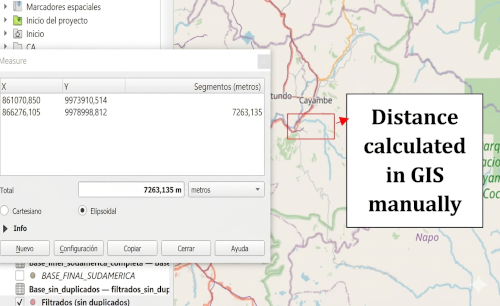}}\label{dist1}}
\hfil
\subfloat[Registered calculated distance.]{\fbox{\includegraphics[width=7cm, height=4.5cm]{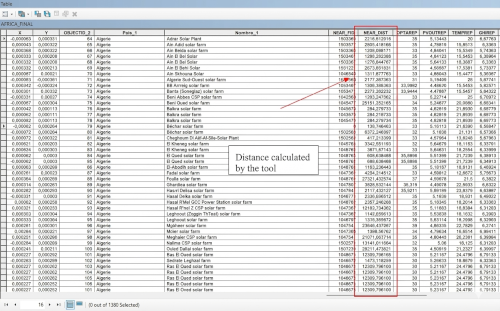}}\label{dist2}}

\subfloat[Area digitized manually.]{\fbox{\includegraphics[width=7cm, height=4.5cm]{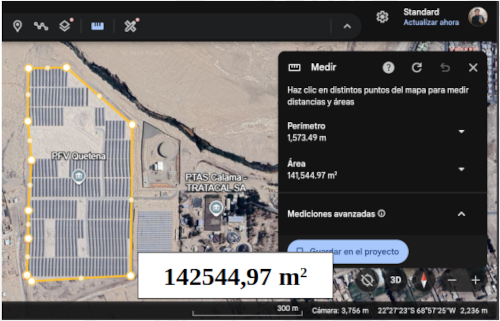}}\label{area1}}
\hfil
\subfloat[Area calculated automatically.]{\fbox{\includegraphics[width=7cm, height=4.5cm]{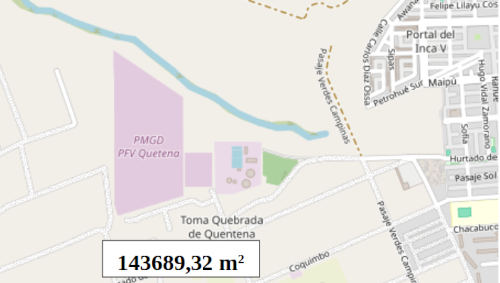}}\label{area2}}
   
	\caption{Calculation of the distance to road and the area of a solar plant.}
	\label{fig:distancia-area}
\end{figure}

Thus, the Euclidean distance from the nearest road point to a solar plant was calculated.
This result is added directly to the dataset's attribute table.
During this process, the calculation of terrain variables was limited exclusively to the sites located within the coverage area of the generated DEMs.
Consequently, the distance to the road network was calculated in a generic manner; however, by using an interactive web platform that integrates these variables, we were able to validate our results and achieve greater accuracy in calculating road distances (Figures \ref{dist1} and \ref{dist2}).

In addition, calculating the \textit{area} required geolocating each PV plant using its latitude and longitude coordinates in order to generate polygons that would allow for an estimate of its surface area.
Initially, polygons were manually digitized in a GIS environment by visually delineating the areas corresponding to solar installations.
However, this time-consuming and labor-intensive process led to the development of an automated procedure using Python code that, based on geographic location (latitude/longitude) and by querying the OSM API, identified polygons tagged as solar plants and automatically calculated their area using a metric reference system.
The following steps were implemented:

\begin{enumerate}
\item Preparation of the initial data, starting with an Excel file containing the geographic coordinates of the plants, along with information on the country, latitude, and longitude.

\item Geographic classification by mapping countries to continents using a logical dictionary, in order to organize the data by continent.

\item Download geospatial data stored in OpenStreetMap PBF files \cite{openstreetmap} for each continent from GEOFABRIK.

\item Filtering relevant information using the “osmium-tool” utility to extract only those features related to solar power plants from OSM data.

\item Based on the filtered data, geographic data for solar power plants by continent is generated in a GeoJSON file.

\item Geographic filtering (bounding box) to retrieve solar plant data for each country. As a result, a country-specific GeoJSON file of solar power plants is generated.

\item The areas of the polygons are calculated (using geodetic methods) to determine the surface area of each solar panel.

\item A spatial join is performed to link the points in Excel (plant locations) with the solar plant polygons obtained from OSM.

\item Finally, an Excel file is generated containing the consolidated information, including the calculated area and the status of each plant.
\end{enumerate}

Automating this calculation made it possible to systematically determine the areas of the polygons for the rest of the world and, at the same time, served as a means of verifying the values previously obtained using the manual method, since there should be no significant differences between the two procedures (Figures \ref{area1} and \ref{area2}).

Finally, the classification of the area is based on industry standards that link the scale of the project to its economic viability and operational purpose \cite{xpert2024tamano}.
Small-scale installations designed for residential and commercial self-consumption typically cover areas of less than 2 hectares ($< 20,000 m^2$), whereas medium-sized solar farms range from 2 to 5 hectares ($20,000$ to $50,000 m^2$), while larger, high-profitability projects cover larger areas, exceeding 5 hectares ($>$50,000 per $m^2$).
After analyzing the specific distribution of our dataset and conducting a visual survey of the plants using \textit{Google Earth}, we decided to make a technical adjustment to these thresholds.
This change responds to the need to more accurately reflect the observed dimensional reality, where the scale of medium and large projects exceeds conventional theoretical standards.
Consequently, the following definitive ranges were established: small ($< 20.000$ $m^2$), medium ($20.000$ $m^2$ to $200.000$ $m^2$) and large ($> 200.000$ $m^2$).

\subsubsection{Climate data}

\begin{figure}[!htbp]
	\centering
	\fbox{\includegraphics[width=0.8\linewidth]{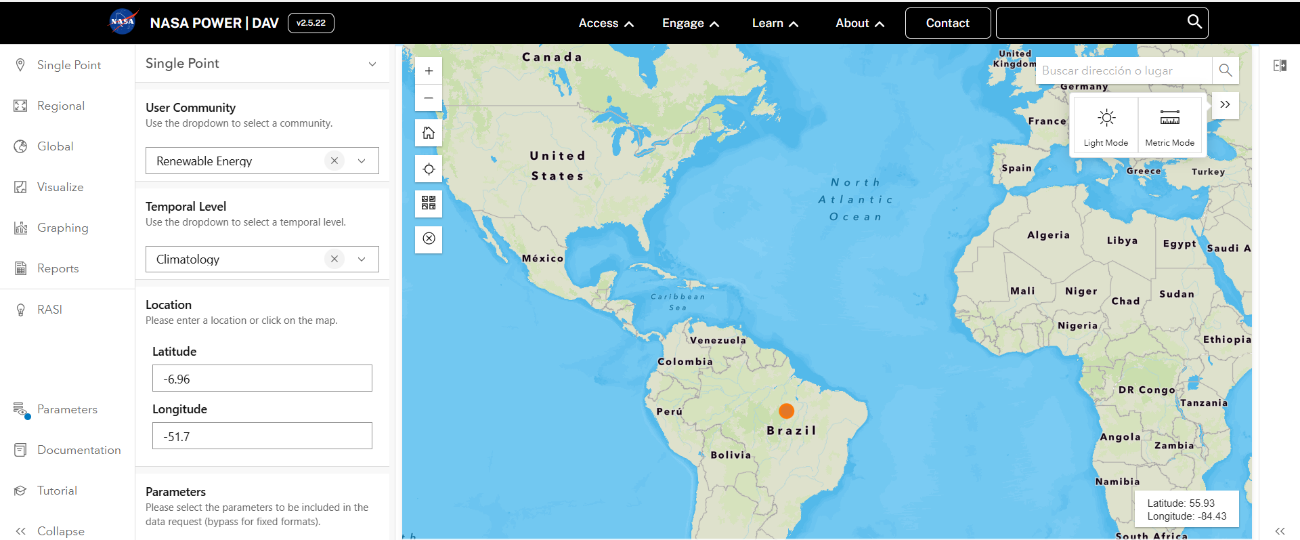}}
	\caption{NASA POWER website.}
	\label{fig:nasapower}
\end{figure}

This type of data is important because it allows us to assess the environmental conditions that affect the performance of solar installations.
The variables considered are \textit{temperature}, \textit{humidity}, \textit{wind speed}, and \textit{wind direction}.
These data were obtained from NASA's Prediction of Worldwide Energy Resources (POWER) \cite{nasa_power}, a NASA initiative of high scientific quality that provides free access to meteorological and solar resource data.
The web interface of this platform (Figure \ref{fig:nasapower}) features an interactive viewer called "POWER Data Access Viewer" (DAV), which allows users to select a specific geographic location (latitude/longitude) or a region, specify a time period and variables of interest, view the data on maps or in graphs, and download the data in CSV format.

Initially, data collection was carried out by manually extracting the necessary parameters from the annual averages.
However, in order to optimize the process and ensure continuous spatial coverage without loss of information, the analysis was scaled up to utilize data from the ERA5 reanalysis \cite{era5}, available in the Climate Data Store (CDS) of the European Union’s Copernicus program.
Using the “Datasets” option, we downloaded “ERA5 monthly averaged data on single levels,” which provides global monthly averages of atmospheric variables with continuous spatial coverage and temporal consistency.

This change in methodology made it possible to transition from individual queries to bulk downloads of global monthly averages in NetCDF (Network Common Data Format), using the subregion extraction option and defining global limits of latitude (-$90\degree$ to $90\degree$) and longitude (-$180\degree$ to $180\degree$) to ensure complete coverage of the study area and avoid loss of spatial information.
After importing these files into QGIS, the base variables (air temperature at 2 m, dew point temperature at 2 m, and the zonal and meridional wind components at 10 m [u10 and v10]) were processed, and relative humidity, wind speed, and wind direction were calculated using raster operations (map algebra) in QGIS. Previously, monthly averages were calculated to obtain representative annual values.

Finally, based on wind direction, the categorical variable \textit{DT\_wind} was created, which was classified into eight categories corresponding to the cardinal directions and the main intermediate points (N, NE, E, SE, S, SW, W, NW), each representing a 45° sector on the compass.

\subsubsection{Power data}

This section examines the \textit{status} of the solar project, i.e., whether it is operational, under construction, proposed, canceled, or on hold.
This data is obtained directly from a CSV file of solar plants worldwide downloaded from Global Energy Monitor.

A key contribution of this study is the calculation of \textit{solar aptitude}, which is derived from previous variables.
This indicator is of great importance for assessing the suitability of an area for the installation of solar energy systems.
To calculate the solar aptitude index ($I_{AS}$), raster layers of terrain variables are used: slope, aspect, shading, and terrain curvature.
These layers are imported into the GIS and standardized to the same reference system and spatial resolution. They are normalized to a common scale, enabling the comparison and combination of heterogeneous variables.
Variables can be assigned different weights based on their relative importance in the installation of solar systems, using methods such as AHP (Analytic Hierarchy Process) or expert judgment.
Standardized and weighted layers are integrated using map algebra, by applying a weighted sum or superposition operation (Eq.\ref{ec1}).

\begin{equation}
I_{AS} = (Slp * 0.40) + (Asp * 0.25) + (Hill * 0.20) + (Curv * 0.15)
\label{ec1}
\end{equation}

The calculated index is classified into suitability categories (low, medium, and high), resulting in a new variable called \textit{T\_aptitude}.
This makes it easier to interpret spatial data, especially when viewing the solar suitability map, which is a tool designed to identify and classify favorable geographic areas based on their potential for harnessing solar energy.
The classification of the solar suitability index into low (0–0.4), medium (0.4–0.6), and high (0.6–1) categories is based on the standardization of results typical of multi-criteria evaluation methods such as AHP \cite{Saaty1980}.
This approach allows results to be expressed on a scale from 0 to 1 and classified into qualitative categories to facilitate their interpretation.
The territory is classified into zones of varying suitability, making it possible to objectively identify the sites with the best conditions for harnessing solar energy, support energy planning, and optimize decision-making for photovoltaic or solar thermal projects.

The \textit{optimal tilt} variable corresponds to the mounting angle that maximizes the output of a fixed PV system oriented toward the equator.
In the Solar Atlas, this parameter is estimated using algorithms derived from the Solargis solar model, which evaluate the system's energy output for different tilt angles.
Next, the angle that provides the highest energy efficiency over the 2018–2024 period is selected, based on the information available on the official website.
This information is represented in raster images, which are imported into the GIS, and the pixel value containing the optimal tilt data is manually extracted.

The \textit{global irradiation} is obtained from a GSA raster by selecting the area of interest.
This platform provides various metrics, such as direct normal irradiance (DNI), horizontal global irradiance (GHI), horizontal diffuse irradiance (DHI), and global irradiance on an inclined plane (GTI).
For this study, we used this last variable, which was overlaid onto the geographic layer of the solar plants using an “Extract Values to Point” operation, allowing us to assign each plant the irradiance value corresponding to its location.

The \textit{capacity} of PV plants is listed in the GEM inventories.
The registered value is obtained through a systematic process of collecting and validating public data.
Researchers compile information from official sources (government data, permits, and reports from power generation companies), industry reports, and verified media reports, which are then cross-checked against other public and private databases to ensure consistency and quality.
Each documented solar plant has a profile page with detailed references \cite{gem_wiki}, and all rated capacities reflect the declared power of the generating units under standard conditions.

Finally, the \textit{PV potential} represents the estimated specific electricity generation for a standard photovoltaic system.
The corresponding raster from the Solar Atlas is downloaded, generated using data on solar radiation, panel efficiency, and climate variables. This raster is integrated with the solar plant layer through a spatial overlay operation, which makes it possible to extract and assign a potential value to each installation.
This value is useful for feasibility studies and energy planning at the regional and national levels.

\subsection{Data integration}

This stage is crucial for consolidating diverse information into a single coherent and consistent structure. This integration process required the unification of variables from different data categories: identification, location, topography, logistics, climate, and operational parameters.

One of the main challenges was to verify that the location points corresponded to actual solar installations.
To this end, a cross-validation was performed between GSA and GEM, using ArcGIS and QGIS to overlay the points onto irradiance layers.
This process was useful to identify inconsistencies, such as incorrect coordinates or sites located in areas unsuitable for PV generation. As a result, it ensured that only actual, correctly georeferenced solar plants were included.

Another challenge was to precisely delineate each solar installation, which is useful for calculating its respective area.
First, the spatial and geometric verification of the location was carried out using ArcGIS satellite maps as a cartographic reference.
The contour of each plant was then digitized manually (polygonization) using an appropriate zoom level that allowed the geometries to be delineated with sufficient precision.
The area was then calculated using ArcMap's geometric calculator.
The area and distance-to-road variables were validated through manual measurements in the GIS, using a random sample of 100 cases to verify the accuracy of the values obtained.

During the digitization phase, multiple records were identified for the same physical location of solar panels, which resulted in redundancy.
For example, in the case of Brazil, a solar installation appeared four times with identical geographic coordinates and descriptive parameters.
The identification of this spatial issue was based on the matching of latitude and longitude attributes.
Data cleaning was performed using the attribute-based duplicate detection.
In addition, spatial outliers outside the coverage area of the rasters were removed using distribution metrics such as "Average Nearest Neighbor".

Once the data had been cleaned, the coordinate system was standardized to WGS 84 by reprojecting the original layers, ensuring spatial compatibility.
In addition, the spatial resolution was standardized to 9 arc-seconds.
Since the solar resource layers (global, direct, and diffuse irradiance) have this nominal resolution, the complementary layers (photovoltaic potential, terrain elevation, and climate variables) were resampled to match that resolution.

As a complementary step, the solar aptitude map was validated to ensure its spatial coherence and consistency with solar radiation patterns and terrain characteristics.
To this end, a comparison was made with the locations of existing solar power plants, revealing that they are concentrated in areas with high solar aptitude.

The actions described above enabled us to compile a single, structured, and consistent dataset with high spatial accuracy and high data quality, which better reflects the characteristics of solar power plants.
This result provides a solid foundation for its subsequent use in statistical analysis.

\section{Statistical Analysis of the Dataset}
\label{sec:stats}

This section presents a statistical characterization of the geographic, topographic, climatic, logistics, and energy-related variables in the created dataset.
Identification variables (code and name) are excluded, as these attributes serve only as labels to distinguish each PV plant.
In addition, the numerical variables of area, slope, curvature, and aspect were used to generate the corresponding categorical variables: size, slope type, curvature type, and aspect type.
This enables more meaningful and understandable statistical comparisons.
The statistical analysis is based on descriptive plots and indicators generated using the R programming language within the RStudio integrated development environment.

\begin{figure}[htbp]
	\centering

\subfloat[Country barplot: clear leadership in solar photovoltaic infrastructure by certain nations.]{\includegraphics[width=0.45\linewidth]{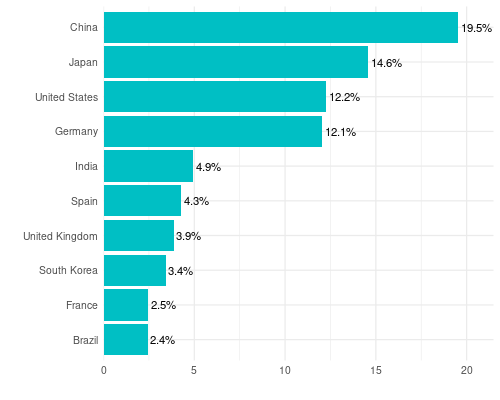}\label{pais}}
\hfill
\subfloat[Continent: the pattern shown in the previous diagram (a) is also reflected at the continental level.]{\includegraphics[width=0.45\linewidth]{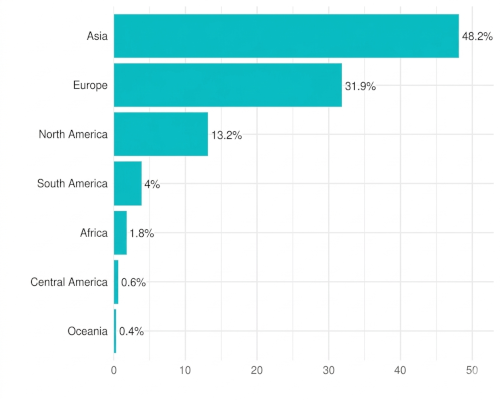}\label{continente}}

\subfloat[Latitude histogram: min: $-41.61$; max: $59.94$; mean: $34.86$; median: $37.01$; std. dev.: $16.01$.]{\includegraphics[width=0.45\linewidth]{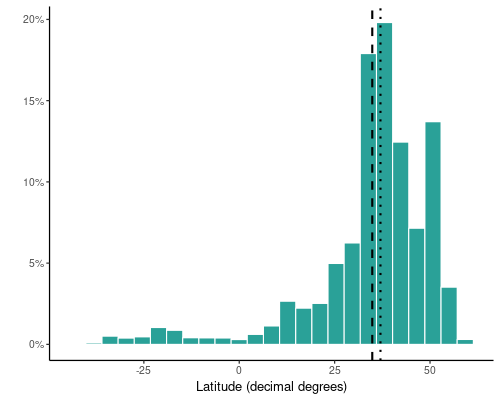}\label{latitude}}
\hfill
\subfloat[Longitude: min: $-124.01$; max: $177.98$; mean: $43.68$; median: $26.76$; std. dev.: $77.31$.]{\includegraphics[width=0.45\linewidth]{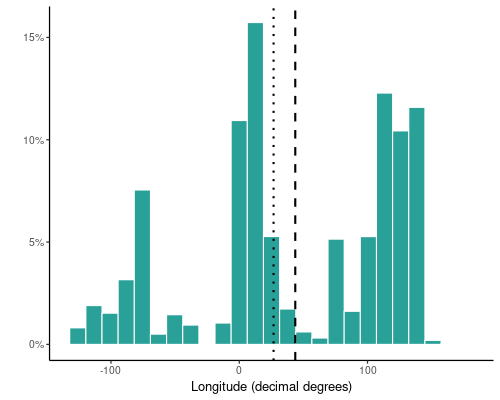}\label{longitude}}

\subfloat[Elevation: min: $-378$; max: $5656$; mean: $353.62$; median: $156$; std. dev.: $526.64$.]{\includegraphics[width=0.45\linewidth]{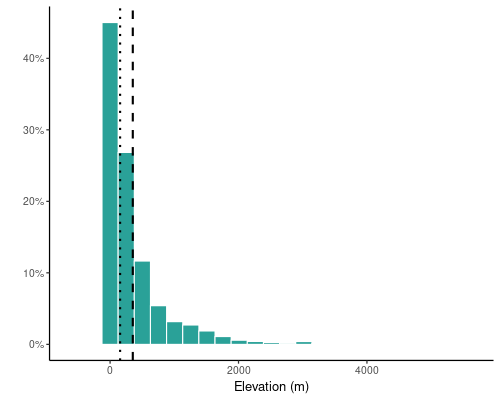}\label{elevation}}
\hfill
\subfloat[Slope: min: $0$; max: $34.68$; mean: $1.45$; median: $0.67$; std. dev.: $2.26$.]{\includegraphics[width=0.45\linewidth]{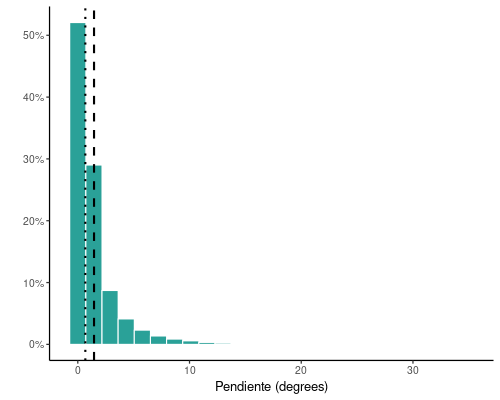}\label{slope2}}

\phantomcaption
\end{figure}

\begin{figure}
\ContinuedFloat
    \centering

\subfloat[Slope type: a strong preference for flat or slightly sloping terrain for the installation of solar panels.]{\includegraphics[width=0.45\linewidth]{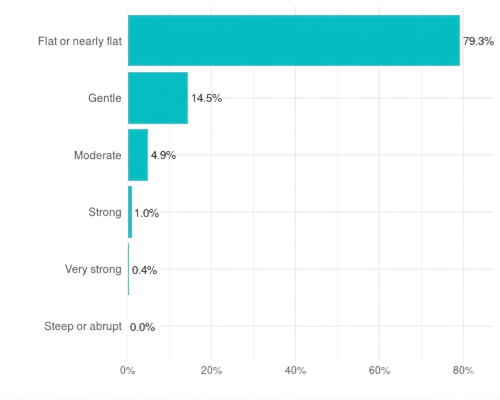}\label{slope}}
\hfill
\subfloat[Aspect: min: $-1.00$; max: $359.61$; mean: $171.59$; median: $170.67$; std. dev.: $98.54$.]{\includegraphics[width=0.45\linewidth]{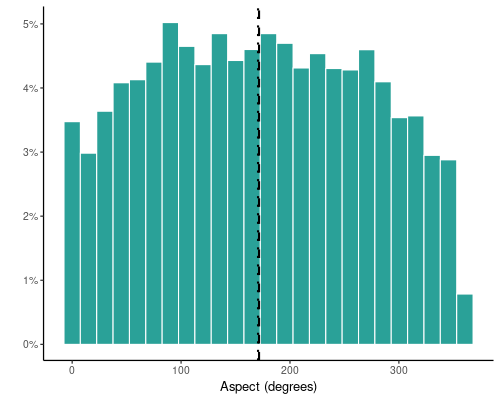}\label{aspect2}}
\hfill

\subfloat[Aspect type: predominantly plants on slopes facing south, east, and southeast.]{\includegraphics[width=0.45\linewidth]{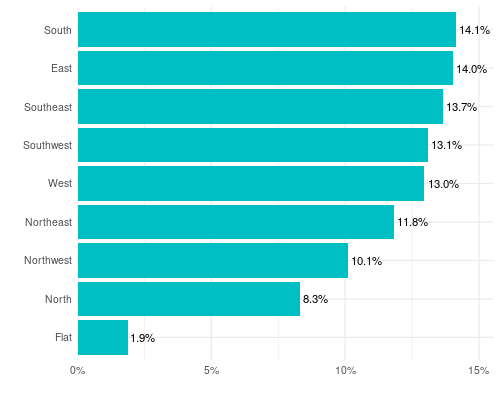}\label{aspect}}
\hfill
\subfloat[Curvature: min: $-0.17$; max: $0.27$; mean: $-0.0006$; median: $0.0001$; std. dev.: $0.014$.]{\includegraphics[width=0.45\linewidth]{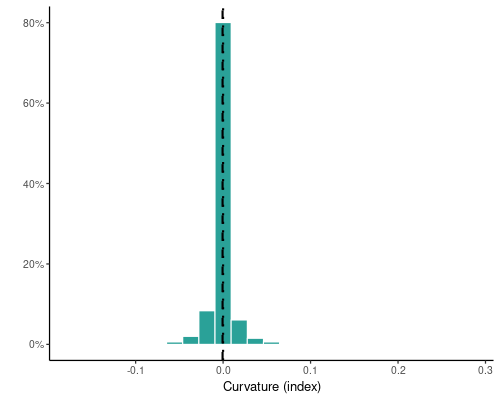}\label{curvature2}} 

\subfloat[Curvature type: a clear preference for terrain with little or no curvature.]{\includegraphics[width=0.45\linewidth]{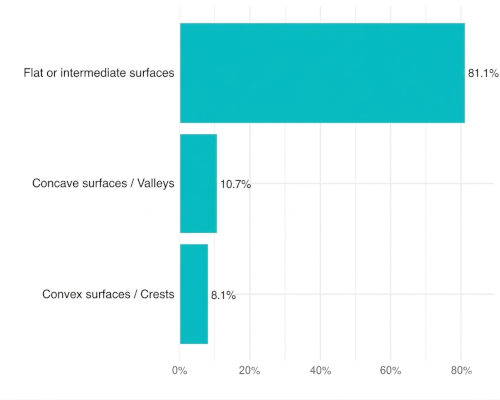}\label{curvature}}
\hfill
\subfloat[Distance to road: min: $0.00$; max: $6309.67$; mean: $1131.61$; median: $562.41$; std. dev.: $1412.82$.]{\includegraphics[width=0.45\linewidth]{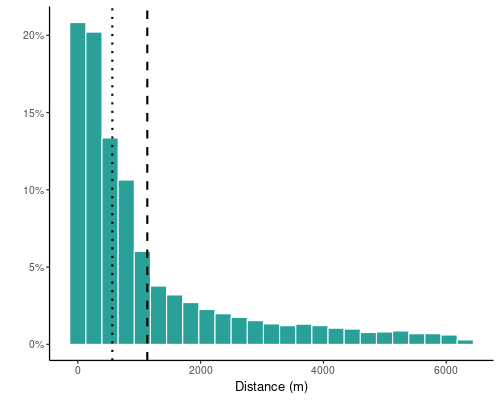}\label{road}}

\phantomcaption
\end{figure}

\begin{figure}
\ContinuedFloat
    \centering

\subfloat[Area: min: $0.013$; max: $153488$; mean: $22641.16$; median: $2813.24$; std. dev.: $34540.73$.]{\includegraphics[width=0.45\linewidth]
{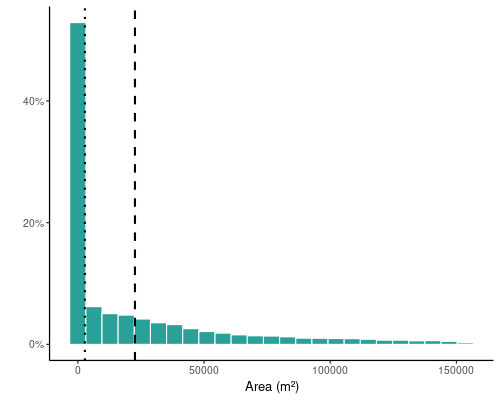}\label{area}}
\hfill
\subfloat[Size: indicates a global trend toward small-scale solar power plants.]{\includegraphics[width=0.45\linewidth]{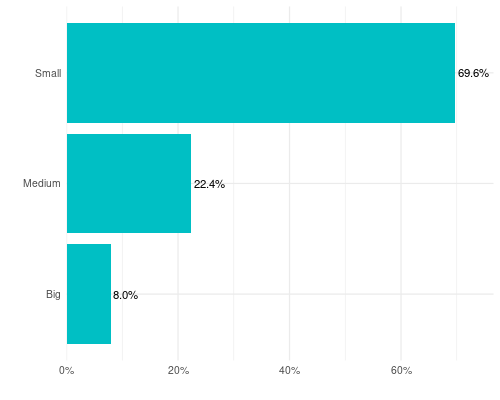}\label{tamano}}

\subfloat[Temperature: min: $-17.06$; max: $30.34$; mean: $14.59$; median: $13.74$; std. dev.: $5.77$.]{\includegraphics[width=0.45\linewidth]{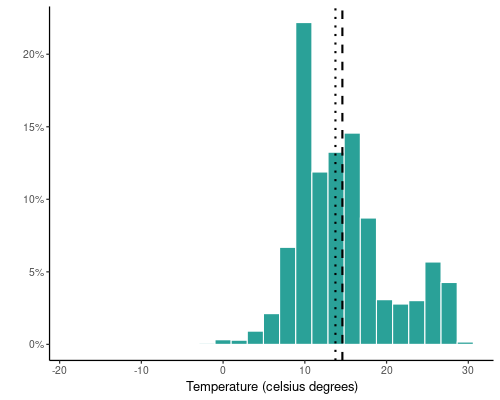}\label{temperature}}
\hfill
\subfloat[Humidity: min: $0$; max: $95$; mean: $66.89$; median: $70.25$; std. dev.: $11.64$.]{\includegraphics[width=0.45\linewidth]{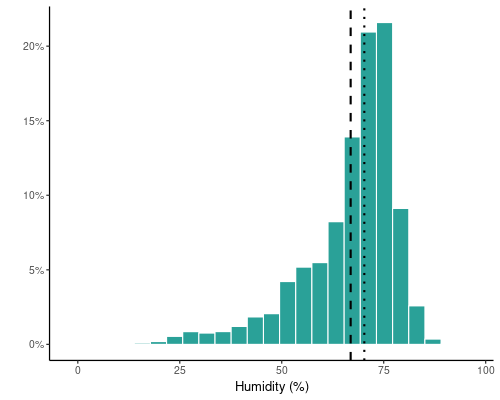}\label{humidity}}

\subfloat[Wind speed: min: $0.00$; max: $12.75$; mean: $0.99$; median: $0.78$; std. dev.: $0.93$.]{\includegraphics[width=0.45\linewidth]{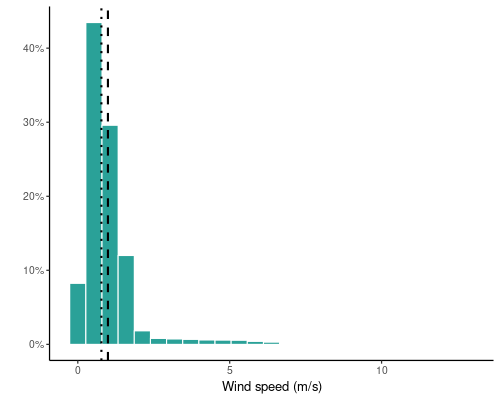}\label{winds}}
\hfill
\subfloat[Wind direction: min: $0.00$; max: $352.50$; mean: $192.97$; median: $218.60$; std. dev.: $59.19$.]{\includegraphics[width=0.45\linewidth]{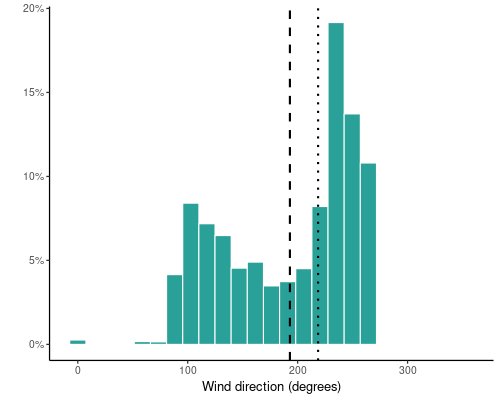}\label{windd2}}

\phantomcaption
\end{figure}

\begin{figure}
\ContinuedFloat
    \centering

\subfloat[Wind direction categories: predominantly southerly winds combined with easterly and westerly winds.]{\includegraphics[width=0.45\linewidth]{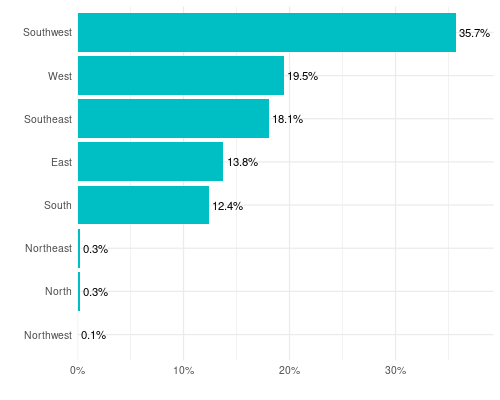}\label{windd}}
\hfill
\subfloat[Operational status: the vast majority of solar power plants are currently generating electricity.]{\includegraphics[width=0.45\linewidth]{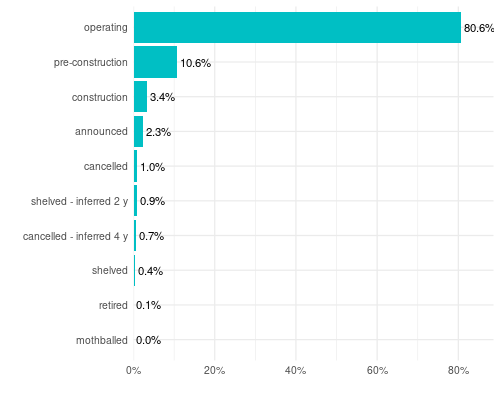}\label{status}}

\subfloat[Solar aptitude: min: $0.00$; max: $0.94$; mean: $0.68$; median: $0.69$; std. dev.: $0.13$.]{\includegraphics[width=0.45\linewidth]{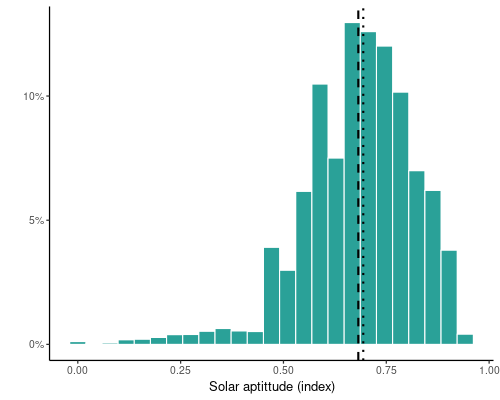}\label{aptsolar2}}
\hfill
\subfloat[Solar aptitude type: a preference for terrain with high capacity for efficient energy generation.]{\includegraphics[width=0.45\linewidth]{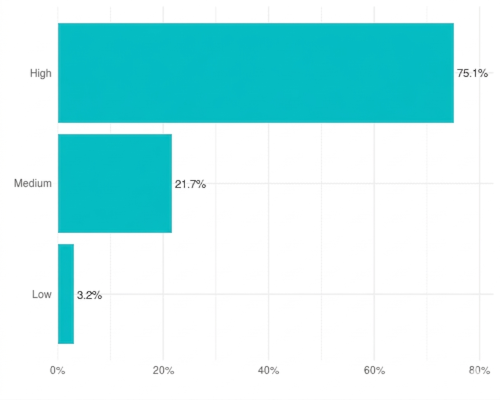}\label{aptsolar}}

\subfloat[Optimal tilt: min: $-2.01$; max: $49$; mean: $31.27$; median: $33.57$; std. dev.: $7.51$.]{\includegraphics[width=0.45\linewidth]{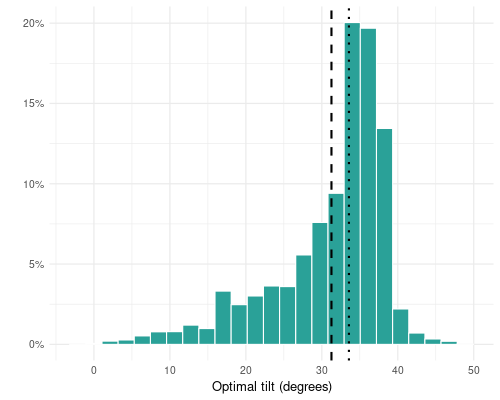}\label{tilt}}
\hfill
\subfloat[Global solar irradiation: min: $0.00$; max: $8.05$; mean: $4.55$; median: $4.47$; std. dev.: $0.86$.]{\includegraphics[width=0.45\linewidth]{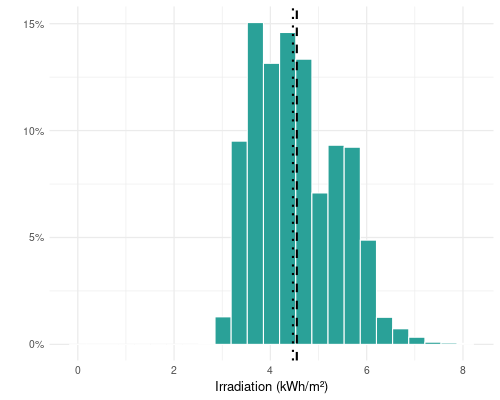}\label{ghi}}

\phantomcaption
\end{figure}

\begin{figure}
\ContinuedFloat
    \centering

\subfloat[Capacity: min: $1.00$; max: $57.50$; mean: $9.32$; median: $3.1$; std. dev.: $13.18$.]{\includegraphics[width=0.45\linewidth]{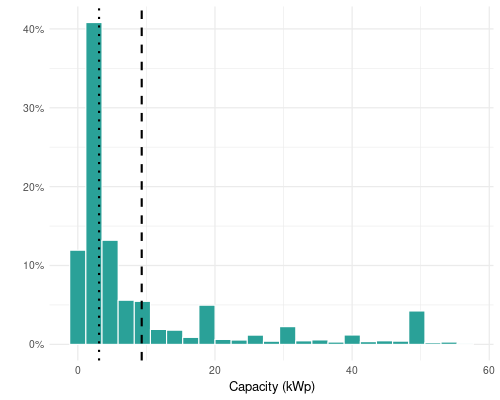}\label{capacity}}
\hfill
\subfloat[PV potential: min: $0.00$; max: $6.39$; mean: $3.80$; median: $3.78$; std. dev.: $0.66$.]{\includegraphics[width=0.45\linewidth]{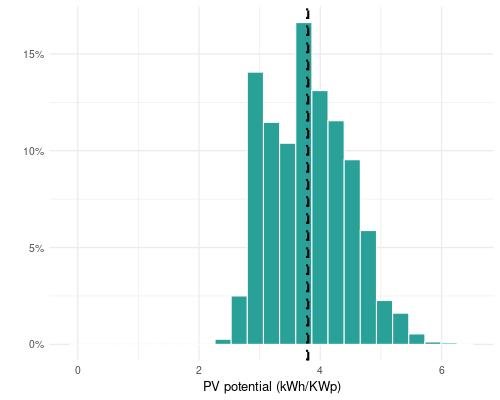}\label{pv}}

\caption{Distribution of the values of the variables in the dataset.}
\label{fig:distributions}
\end{figure}

\subsection{Geographical variables}
It is essential to know where they are located and to understand the geographic distribution of solar power plants.
Globally, a small number of countries account for a large proportion of solar power plants.
China tops the list with nearly a quarter of all plants, while Japan, the United States, and Germany also account for a significant share (Figure \ref{pais}).
This trend is mirrored on a continental scale, where Asia, Europe, and North America are the leading regions in terms of PV solar power (Figure \ref{continente}).
The distribution of longitude is clearly multimodal (Figure \ref{longitude}), identifying three specific regions for PV development: the European axis (0° to 30° E), the Asian axis (100° to 150° E), and the North American axis (-120° to -70° W). 
In contrast, latitude (Figure \ref{latitude}) reveals a massive concentration of data in the Northern Hemisphere. With an average of 34.86°N, most plants are located in the industrial “Solar Belt” (China, the United States, and Southern Europe).

\subsection{Topographical variables}
These variables enable an analysis of the soil or terrain conditions where the solar infrastructure is located.
The elevation data show that more than 50\% of solar power plants are located in lowland areas (Figure \ref{elevation}), with an average elevation of 353.61 meters above sea level, which is advantageous from a logistical and economic standpoint (proximity to cities and to electrical and road infrastructure).
The vast majority of solar power plants (approximately 95\%) are located on land with little or no slope, falling into the categories of flat, nearly flat, and slightly sloped (Figures \ref{slope2} and \ref{slope}).
This makes sense both technically and economically, as it facilitates installation, simplifies operation and maintenance, reduces costs, and improves efficiency. The maximum value (34.68°) sets the upper limit of technological adaptability for plants on steep terrain.
With regard to aspect or orientation (the direction the land faces), the average value is $171.58^{\circ}$ (Figure \ref{aspect2}), very close to $180^{\circ}$.
The majority of installations are located on slopes facing south, east, and southeast, accounting for more than 40\% (Figure \ref{aspect}).
This distribution confirms that most plants are located in the Northern Hemisphere, with their leaves oriented toward the equator.
More than 80\% of the data show a curvature close to 0, indicating a clear preference for areas with low or no curvature (Figures \ref{curvature2} and \ref{curvature}), where the terrain is geomorphologically stable and conducive to the efficient development of PV infrastructure.

\subsection{Logistics variables}
These variables are what can determine the feasibility of building a solar energy project.
A key logistic metric is the distance to the road (Figure \ref{road}).
Nearly 60\% of solar power plants are located less than 1 km from an existing road, including those immediately adjacent to the road network (minimum of 0).
This demonstrates that access to transportation infrastructure is a critical factor, so it is important to select sites that already have transportation links for personnel, panels, and heavy machinery.
There are also plants located more than 6 km from the main road lines.
These are typically megaprojects in desert or remote areas, where the abundant solar resources justify the massive investment.
One of the factors that influences the planning of resources, transportation, and implementation of a solar project is the size of the plant.
In solar engineering, size determines operational complexity.
As for the size of solar power plants (Figure \ref{area}), there is considerable variation, with thousands of small PV plants coexisting alongside a few “giants” that generate the majority of the energy.
More than 50\% of the plants are concentrated in the range of 0 to 10,000 $m^2$, suggesting a significant presence of industrial and commercial self-consumption facilities or small distributed generation parks.
The above dimensions were grouped into size categories. There are a large number of small plants, a moderate number of medium-sized plants, and a few large plants (Figure \ref{tamano}). 
This is consistent with the global trend toward distributed solar power generation; that is, the sharp increase in residential and commercial installations, in contrast to large-scale projects, which require greater investment and are more complex.

\subsection{Climate variables}

Regarding the climatic conditions at the sites of solar power plants, one of the most critical climatic variables is temperature (Figure \ref{temperature}).
This variable represents a loss factor: as the cell temperature rises, the panel's efficiency decreases.
The data distribution shows a clear predominance in temperate and warm climates, but the average of $14.59 \text{ °C}$ indicates that most plants operate under conditions where the air helps keep the panels at temperatures ideal for performance.
Most plants operate in areas with moderate to high humidity levels (Figure \ref{humidity}), with an average of $66.90\%$.
Plant design must take into account corrosion processes in metal structures and material degradation caused by the constant presence of water vapor in the atmosphere.
Wind acts as a thermal regulator and a structural factor.
The wind speed (Figure \ref{winds}) indicates a predominantly low-turbulence operating environment, with an average of $0.99 \text{ m/s}$.
Although this characteristic is somewhat detrimental to panel cooling and efficiency, it is highly beneficial for the long-term structural stability of PV arrays.
The wind direction distribution is clearly bimodal (Figure \ref{windd2}), with two main wind directions for most solar plants.
A large concentration of plants is exposed to winds coming from between 220° and 260° (southwest), while a secondary area is exposed to winds coming from between 100° and 130° (southeast).
A strategic correlation is observed between wind direction (Figure \ref{windd}) and the orientation of solar plants.
Winds blowing in a certain direction are most effective at cooling panels facing that same direction, since they strike the front surface of the panels directly.

\subsection{Power variables}
These variables enable analysis of how solar photovoltaic plants are designed and how much energy they generate.
The operational status is related to the life cycle of solar power plants (Figure \ref{status}).
The vast majority of plants are currently in operation; a smaller proportion consists of projects in the pre-construction and construction phases, and there is a low incidence of failed or withdrawn projects.
This is a positive development, as it reflects maturity in electricity generation and the ongoing expansion of solar capacity.
As for how “suitable” a site is for power generation, most values fall between 0.6 and 0.85 (Figure \ref{aptsolar2}), confirming that these sites are not only “suitable” but are very close to their theoretical maximum potential.
This suggests that site selection criteria have prioritized areas with maximum theoretical efficiency.
Approximately 75\% of solar plants fall into the “High” solar suitability category (Figure \ref{aptsolar}).
This shows that the global solar industry does not install systems at random; sites are selected by prioritizing locations where solar energy is naturally optimal.
Installation in an area with moderate or low suitability is typically driven by specific reasons, critical needs, or regions with high energy demand.
In the context of a global dataset, these are outliers or atypical cases, whose presence adds diversity, allowing us to analyze different realities and identify patterns of efficiency outside of ideal conditions.
The optimal tilt angle corresponds to 45\% of the plants falling within the range of 30° to 40°, with an average of 31.27° (Figure \ref{tilt}).
This confirms that the design follows a rule of thumb for maximizing energy efficiency: the optimal tilt angle is typically approximately equal to the latitude of the location (mean of 34.86°).
Irradiance is an indicator of available solar energy (Figure \ref{ghi}).
Their values are not scattered randomly; rather, they cluster around radiation values known to be optimal for the industry (similar to a Gaussian bell curve), but with a bias toward locations with high solar resources.
The capacity diagram shows a high degree of heterogeneity (Figure \ref{capacity}), demonstrating that it covers the entire spectrum of the photovoltaic industry.
However, nearly 60\% of the plants have an installed capacity of less than 10 MW.
This suggests that medium-sized solar farms and distributed generation projects have reached maturity.
Finally, PV potential (kWh/kWp) is the most important performance metric; it tells us how much energy can actually be converted into electricity (Figure \ref{pv}).
The vast majority of plants fall within the high-efficiency range ($3.5$ to $4.2 \text{ kWh/kWp}$), with an average of $3.80 \text{ kWh/kWp}$ and a low standard deviation ($0.65$).
This demonstrates that the global infrastructure is located in areas of high energy productivity.

\section{Conclusions}
\label{sec:conclusion}

This study has resulted in the creation of one of the most comprehensive and multidimensional datasets on photovoltaic infrastructure worldwide to date.
It consists of nearly 59,000 records of solar power plants, each with 27 attributes related to geography, topography, logistics, climate, and power, obtained from multiple sources.
This expands the scope of existing solar energy datasets.
Given its coverage, level of detail, and reproducible methodology, the dataset represents a reliable and complementary alternative to the solar data resources currently available.

A thorough review and analysis of scientific literature and available data platforms provide an objective, evidence-based foundation for attribute selection, thereby avoiding the inclusion of irrelevant parameters.
Unlike current datasets, this study incorporates critical variables such as the area and distance from the road for each of the thousands of solar plants recorded.
Once the attributes have been defined, it is advisable to organize them in a clear and consistent manner.
The dataset architecture was based on a taxonomy comprising several key dimensions, which provides a more comprehensive view of the photovoltaic infrastructure.
Subsequently, gathering the data required not only compiling information from various reliable sources, but also processing that information and generating new insights to fill in the gaps in the sources.
Consolidating this diverse set of data requires a rigorous process of cleaning and verification to ensure its proper integration into a single dataset.
The workflow described here constitutes a systematic and reproducible methodology that can be expanded or adapted by future researchers in the field of renewable energy.

The combination of different categories of variables is not typically found in a single source of solar data. 
This represents a significant improvement over the fragmented and incomplete datasets available in both the public and private sectors.
This integrated approach provides a more comprehensive and multidimensional characterization of existing solar photovoltaic plants.
As a result, the dataset created serves as a valuable tool for future research and studies on the optimal siting of solar installations, strategic planning, data-driven modeling and forecasting, geospatial analysis and simulation, and artificial intelligence applications, among many other uses.

The integration of GIS and Python enabled the implementation of automated processes, which were essential not only for consolidating the nearly 59,000 records that make up the dataset, but also for generating variables whose derivation required highly complex geospatial calculations.
This study demonstrates that the issues of fragmentation and bias in previous datasets, which were technically intractable using conventional methods, can be overcome through automation and engineering of variables.

The dataset's considerable size gives it enormous statistical power.
An analysis of the 58,978 records revealed several significant patterns.
Operational plants are in the majority, indicating that the dataset primarily represents infrastructure that is currently in operation.
A notable geographical concentration was identified in certain countries, such as China and the United States, as well as in regions of Europe that have limited solar resources but a high level of technological expertise.
Small-scale plants predominate, indicating a clear trend regarding the development of this infrastructure.
The combination of a low slope and zero curvature was also identified as the industry standard, as it minimizes the costs of site preparation and infrastructure construction.
There is a technical and logistical balance involved in selecting a site for a plant that depends not only on solar resources, but also on proximity to road networks and the management of large land areas.
There is a strategic correlation between wind direction and the orientation of the panels, which enhances cooling and energy efficiency.
The deployment of solar power plants demonstrates rigorous energy efficiency in their design, where the latitude and the optimal tilt of the panels are closely related, resulting in a robust specific photovoltaic output.
These results confirm that solar energy is a mature, predictable, and highly productive technology in various regions around the world.

As a future task, it would be advisable to expand the range of variables in the dataset, incorporate mechanisms for periodic updates, and conduct a more in-depth validation of derived attributes to further enhance its usefulness in research and practical applications.

\section{Acknowledgments}
The authors would like to thank the support team formed by young students from the Central University of Ecuador, whose contributions were essential to the completion of this work.

\section{Availability and Access to the Dataset}
The data and code supporting the findings of this study are available on GitHub (https://github.com/cimejia/solarPV).

\bibliographystyle{unsrt}
\bibliography{references}  


\end{document}